\documentclass[doublespacing]{elsart}
\usepackage{graphicx,harvard,amssymb}
\journal{New Astronomy}

\newcommand{\lsim}{\,\lower2truept\hbox{${<\atop\hbox{\raise4truept\hbox{$\sim$}}}$}\,}
\newcommand{\gsim}{\,\lower2truept\hbox{${>\atop\hbox{\raise4truept\hbox{$\sim$}}}$}\,}

\def\hi{H$_{\rm I}\,$}
\def\hii{H$_{\rm II}\,$}

\def\he{He$_{\rm I}\,$}
\def\hee{He$_{\rm II}\,$}
\def\heee{He$_{\rm III}\,$}

\begin{document}

\begin{frontmatter}

\title{Radiative effects by high-z UV radiation background:
Implications for the future CMB polarization measurements}

\author[Popaqui,Popali]{L.A.~Popa},
\author[Popaqui]{C.~Burigana},
\author[Popaqui]{N.~Mandolesi},
\vskip 0.5truecm
\address[Popaqui]{INAF/IASF, Istituto Nazionale di Astrofisica,
Istituto di Astrofisica Spaziale e Fisica
Cosmica, Sezione di Bologna,
Via Gobetti 101, I-40129 Bologna, Italy}
\address[Popali]{Institute for Space Sciences,
Bucharest-Magurele R-76900, Romania}


\footnote{The address to which the proofs have to be sent is: \\
Lucia A. Popa, INAF/IASF, Sezione di Bologna,
Via Gobetti 101, I-40129, Bologna, Italy\\
fax: +39-051-6398724\\
e-mail: popa@bo.iasf.cnr.it}

\newpage
\begin{abstract}

The radiative effects by
high-redshift ultraviolet radiation background (UVB)
heats the gas in the IGM and eliminate the neutral hydrogen and helium
that dominate the cooling of the primordial gas.
We investigate the role of the radiative  effects
for the temporal evolution of the reionization fraction
by using cosmological Smooth Particle Hydrodynamics (SPH)
simulations.\\
We find that the increase of photo-ionization and photo-heating
rates due to optical depth effects results in a
significantly contribute to the heating of the IGM
before and during the reionization.
The main effect of the
UV radiation spectrum
on the temporal evolution of the ionization
fraction is given by the value of the reionization
redshift, $z_{re}$, and the redshift interval,
$\Delta z$, in which the reionization is completed.\\
We evaluate the effects of the UV radiation background
on the CMB angular power spectrum
taking into account different
temporal evolutions of the ionization fraction.
We find that for reionization models
with degenerated CMB temperature anisotropy power spectra,
the radiative  mechanisms leave
distinct signatures on the E-mode polarization power spectrum,
at large scales ($l<50$).
We  show that through E-mode CMB polarization
power spectrum measurements,
the {\sc Planck} experiment will have the sensitivity to distinguish
between
different reionization histories even when
they imply the same optical
depth to electron scattering and
degenerated $C_T$ power spectra.

{\it This work has been done in the framework of the
{\sc Planck} LFI activities.}
\end{abstract}


\end{frontmatter}

\newpage

\section{Introduction}

The detailed study of the intergalactic medium (IGM)
is fundamentally important for understanding the
large-scale structure properties and the galaxy formation process.\\
At the epoch of  reionization the collapsed objects began to influence
the diffuse gas in the IGM and rendered it transparent to the ultraviolet
photons.
In the hierarchical models of structure formation, an early UV radiation
background (UVB) is produced by the first collapsed halos with masses near
 the cosmological Jeans mass.
In order to virialize in the potential wells of the dark matter halos,
the gas in the IGM must have a mass, $M_b$,
greater than the Jeans mass, $M_J$, which
is $\sim 10^5 M_{\odot}$ at a redshift $z\sim 30$,
corresponding to a virial temperature $T_{vir} \sim 10^4$K.
Photoionization by the high-redshift UVB  heats the low density gas in the IGM before
it falls into the dark matter
wells, strongly reducing the fraction of
neutral hydrogen and helium that
dominate the cooling of the primordial gas at temperatures of $T_{vir}$.\\
The primordial gas dynamics in the photoionized IGM
gives important insight into the epoch of the reionization
and the end of the ''dark ages`` of the universe,
believed to have been caused by the light from the first
generation of massive stars (see, e.g., the review by Barkana and Loeb
(2001) and the references therein). \\
From the observational point of view, the study of the
reionization process itself as well as  the properties
of the sources driving it is challenged by a variety of
observational probes.
A powerful  observational probe comes from the Ly$\alpha$
absorption spectra of the high redshift quasars
(Becker et al., 2001; Fan et al., 2002; White et al., 2003;
Fan et al., 2004) showing that all
known quasars with $z>6$ have a complete Gunn-Peterson (GP) trough
and a rapid evolving hydrogen neutral fraction compatible with the final
stage of reionization.
This conclusion  is strengthen by the study of the
proximity effects around high redshift  quasars
(Wyithe and Loeb, 2004; Mesinger and Haiman, 2004).
Also, the high value of the IGM temperature inferred from
the study of Ly$\alpha$ forest at $z \sim 4$
indicates that the hydrogen neutral fraction
changed substantially at $z \leq 9$
(Theuns et al., 2002; Hui and Haiman, 2003).\\
An other powerful observational probe is
represented by the high value of the electron scattering optical depth,
 $\tau_{es}$, inferred from
the Cosmic Microwave Background (CMB) anisotropy
measured  by the WMAP experiment
(Kogut et al., 2003; Spergel et al., 2003; Verde et al., 2003)
which requires reionization to begin at $z>14$.\\
In order to reconcile these observations
 a number of semi-analytical models have been developed,
 the most intriguing possibility
 being the so called ''double reionization`` for which the
 globally-average ionization fraction   decreases  with cosmic time
  over a limited period.
 The resulting reionization histories display a wide range of features
  extended over a long redshift interval.
 Some of these works (Cen, 2002; Wyithe and Loeb, 2004) agree
on the necessity of
a first generation of metal-free stars (Population III)
with heavy mass function
to produce the large observed value of $\tau_{es}$.
Other studies (Ciardi, Ferrara and White, 2003; Somerville and Livio,
2000)
found that the metal enriched stars (Population II) are able to reionize the
universe enough early to produce such high values of $\tau_{es}$.
These works assume that the total volume fraction of the ionized
regions were driven by the ionizing sources located in the dark matter
halos.
Typically, these scenarios
require a number of assumptions on the dynamics of the stellar evolution.
Recent works (Scannapieco et al., 2003; Furlanetto and Loeb, 2004)
show that the transition redshift
at which the star formation switches between the
two modes
must occur over a large redshift interval and
only after the reionization is complete.
In this scenarios, the observables characterizing the global
reionization history of the universe
can be reconstructed with a
large range of input parameters. \\
Other physical mechanisms that can explain the extension of
reionization over a
long redshift interval are radiative,
the most important being the  photo-ionization heating.
A recent work (Furlanetto and Loeb, 2004) critically examined
the plausibility of different  mechanisms,
showing that
double reionization requires a rapid drop in
ionizing emissivity over
a single recombination time that can be obtained
with unusual choices of the physical parameters.\\
Cosmological N-body simulations provide a completely different
approach to model the reionization process
(Ostriker and Gnedin, 1996; Gnedin and Ostriker, 1997;
Gnedin, 2000; Ciardi, Stoehr and White, 2003),
offering the advantage
over the semi-analytical models to be able
to fully account for
the dynamical evolution of the matter contents in the universe.
The main limitation of N-body simulations is related to their resolution
that turns to be critical for the reionization studies.
At redshifts $z \ge 3$
the quasar population declines (Madau, Meiksin and Rees, 1999) and
it could not be able
to reionize the gas in the IGM.
The bulk of the ionizing photons at high redshifts
should be then produced  by an early population of stars
(see, e.g., Haiman, Abel and Rees, 2000, Barkana and Loeb, 2000, Cen,
2002). \\
In the hierarchical theories of structure formation
the low mass objects collapse first
merging then into progressively larger systems.
In this scenario,
small mass objects ($ M \sim 10^5 - 10^7 M_{\odot}$)
are the main contributors at high redshifts
while their role becomes much less important
at low redshifts in comparison with that coming from objects
with higher masses ($M > 10^9M_{\odot}$).
Thus, numerical simulations are subjected to the difficulty to be
able to resolve objects in a large dynamic range
with enough resolution.

A number of works have also considered the implications of different
reionization mechanisms for the CMB angular power spectra
(see e.g., Bruscoli, Ferrara and Scannapieco 2002; Holder et al., 2003; Haiman and Holder, 2003;
Naselsky and Chiang, 2004).\\
Although the  electron optical depth to the Thompson scattering
of the CMB photons, is an important cosmological parameter
(see, e.g., Kaplinghat et al., 2003; Hu and Holder, 2003),
the CMB angular power spectrum contains more information
about the reionization process than the optical depth
integrated over the whole ionization history.

In this paper we study the role of the radiative effects
for the time evolution
of the global ionization fraction at sub-galactic scales
($M\sim 5 \times 10^{8}M_{\odot}$)
by using N-body cosmological hydrodynamical simulations.\\
We evaluate the effects of the UV radiation background amplitude
and shape (in both frequency and redshift)
on the features of the CMB anisotropy and polarization angular power
spectra
and address their detectability by ongoing and future CMB
anisotropy experiments. \\
The paper is organized as follows:
in $\S$ 2, we compute
the UV radiation background  flux  as solution to the cosmological
radiative transfer equation and discuss the radiative mechanisms that can
cause the global ionization fraction to evolve non-monotonically
with the cosmic time.
We present  the various reionization histories
obtained for the considered UVB flux models in $\S$ 3. \\
In $\S$ 4
we compute the CMB angular power spectra,  in both temperature
and polarization, for the various reionization histories
by taking into account the radiative  mechanisms.
The possibility to distinguish among different reionization
scenarios with the forthcoming {\sc Planck} experiment
is also discussed in this section.
We draw our main conclusions in $\S$ 4.

Throughout we assume a background cosmology consistent
with the most recent cosmological measurements (Spergel et al., 2003) with
energy density of $\Omega_m=0.27$ in matter, $\Omega_b=0.044$ in baryons,
$\Omega_{\Lambda}=0.73$ in cosmological constant, a Hubble constant of
$H_0$=72 km s$^{-1}$Mpc$^{-1}$, an {\it rms} amplitude of
$\sigma_8=0.84$ for mass density fluctuations
in a sphere of radius 8h$^{-1}$Mpc, adiabatic initial conditions and
a primordial power spectrum with a power-law scalar spectral index
$n_s=1$.

\section{Building up the UV radiation background}

Reionization is an inhomogeneous process that proceeds in a patchy
way. The radiation output associated with the collapsed halos gradually
builds up a cosmic UVB.\\
At early times,
most of
the gas in the IGM is still neutral
and the cosmological \hii regions around the individual
sources do not overlap. At this early stage the gas in the IGM is
opaque to the ionization photons, causing fluctuations
of both ionization fraction and UVB intensity
from region to region.
At the reionization redshift, $z_{re}$,
the \hii regions surrounding
the individual sources in the IGM overlap.
As shown by numerical simulations
(see, e.g., Gnedin, 2000, Barkana and Loeb, 2000),
at $z_{re}$ the UVB intensity
increases sharply and the baryonic
gas is heated to a temperature $T_{IGM} \simeq 10^4$K.
At smaller redshifts, the quasars contribution to the UVB
leads to  $T_{IGM} \simeq 4 \times 10^4$K (Theuns et al., 2002).

The cumulative UV background flux, $J(\nu_o,z_o)$,
observed at the frequency $\nu_o$
and redshift $z_{o}$,
(in units of 10$^{-21}$ erg cm$^{-2}$ s$^{-1}$ sr$^{-1}$ )
due to photons emitted from redshifts between $z_o$
and an effective emission screen located at $z_{sc}\geq z_o$,
is the solution of the cosmological radiative transfer equation
(Peebles, 1993; Haiman, Rees \& Loeb, 1997):
\begin{equation}
J(\nu_o,z_o)=\frac{c}{4 \pi }\int^{z_{sc}}_{z_o}
e^{-\tau_{eff}(\nu_o,z_o,z)}  \frac{dt}{dz}
\left(\frac{1+z_o} {1+z} \right)^3 j(\nu_z,z)dz,
\end{equation}
where: $j(\nu_z,z)$ is the comoving emission
coefficient
(in units of 10$^{-21}$ erg cm$^{-3}$ s$^{-1}$ sr$^{-1}$)
computed at emission redshift $z$ and photon frequency
$\nu_z=\nu_o(1+z)/(1+z_o)$,
 $\tau_{eff}(\nu_o,z_o,z)$ is
the effective optical depth
at the frequency $\nu_o$
due to the absorption of the residual gas in the IGM
between $z_o$ and $z$ and
$(dt/dz)^{-1}=-H_0(1+z)\sqrt{\Omega_m(1+z)^3+\Omega_{\Lambda}}$
is the line element in our $\Lambda$CDM cosmological model.\\
Above the hydrogen ionization threshold ($\nu^{th}_{H_{I}}=13.6$ eV)
the UV radiation background is processed
due to the absorption of residual gas in the IGM
dominated by neutral hydrogen and helium.
At these frequencies the effective optical depth
is given by (see e.g. Haiman, Rees \& Loeb, 1997):
\begin{eqnarray}
\tau_{eff}(\nu_o,z_o,z)=c\int^z_{z_o}
 \frac{dt}{dz} \kappa(\nu_z,z)dz ,
\hspace{0.3cm}
\kappa(\nu_z,z)=\sum_i \sigma_i(\nu_z) n_i(z)\, ,
\end{eqnarray}
where: $i=(H_{I}\,,He_{I}$) and
$\sigma_i$ and $n_i$
are the cross section and the number
density respectively, corresponding to species $i$.\\
Below  hydrogen ionization threshold,
in the frequency range 11.18-13.6 eV
(enclosing Lyman and Werner bands  of H$_{2}$),
the UV background spectrum is processed due to the
absorption into  H$_{2}$ lines (Ciardi, Ferrara \& Abel, 2000;
Haiman, Abel \& Rees, 2000).
The dissociation of H$_2$ molecules removes photons
from the UVB and decreases the rate
of  dissociation process inside the collapsed halos.
The magnitude of this process depends on the H$_2$ fraction
and the redshift distribution of the sources.
Haiman, Abel \& Rees (2000) found that the intergalactic H$_2$
 has a negligible effect on the UVB spectrum, both because its effective optical depth is small ($\leq$0.1) and because it is photodissociated at early stages.

To compute the spectrum of the UV radiation background
taking into account the radiative transfer effects
as given by  eq.~(1), we assume
that the comoving emission coefficient $j(\nu_z,z)$
has a power-law spectrum with redshift dependent
normalization, $j_{21}(z)$,  and
a free spectral index, $\alpha$, of the  form:
\begin{eqnarray}
j(\nu_z,z)=j_{21}(z)
\left( \frac{\nu_z}{\nu^{th}_{HI}}\right) ^{-\alpha}
\hspace{0.2cm} {\rm for} \hspace{0.2cm}
11.18\, {\rm eV} < h\nu_z < 455.6\,{\rm eV} \,.
\end{eqnarray}
The frequency range corresponds to photons collected
at observed frequencies in 11.18 - 13.6 eV bands (LW),
 observed redshifts in the interval 5 - 50,
arriving from  emission screens uniformly distributed in space,
located between $z_o$ and a maximum
redshift  $z^{max}_{sc}$=200. \\
Similar to the model in Gnedin and Hui (1998),
  $j_{21}(z)$ was parameterized as a function of the reionization
redshift, $z_{re}$, and two free parameters, A and B,  in the form:
\begin{equation}
j_{21}(z)= {\rm A} \left[ 1+\tanh\left({\rm B}\,\frac{z_{re}-z}
{1+z}\right)\right]\,.
\end{equation}
For $z \sim z_{re}$, the redshift dependent normalization,
 $j_{21}(z)$,
 has a value proportional to $A$ and
gradually drops for $z>z_{re}$ in a redshift
interval $\Delta z \sim {\rm B}^{-1}$.
This transition period  is motivated by numerical simulations
(Gnedin and Ostriker, 1997)  showing that
the hydrogen neutral fraction
drops steadily
as the formation rate of
collapsed regions of high density increases in time.
The ionizing intensity climbs gradually until the gas
becomes highly ionized ($H_I \lsim 10^{-4}$),
thus completing the reionization process. \\
Each choice of the parameter vector
${\bf p}=({\rm A},{\rm B},z_{re},\alpha)$
in eq.~(4) defines an UV background flux model.\\
\begin{figure}
\caption{Panel a): Redshift evolution of the
averaged UVB flux
(units of 10$^{-21}$ erg cm$^{-2}$ s$^{-1}$ sr$^{-1}$ )
in LW bands.
Panel b): Redshift evolution of the
averaged effective optical depth
in LW bands assuming a constant H$_2$ fraction
of $2\times 10^{-6}$.
The model parameters are indicated in the Table.}
\begin{center}
\includegraphics[width=15cm]{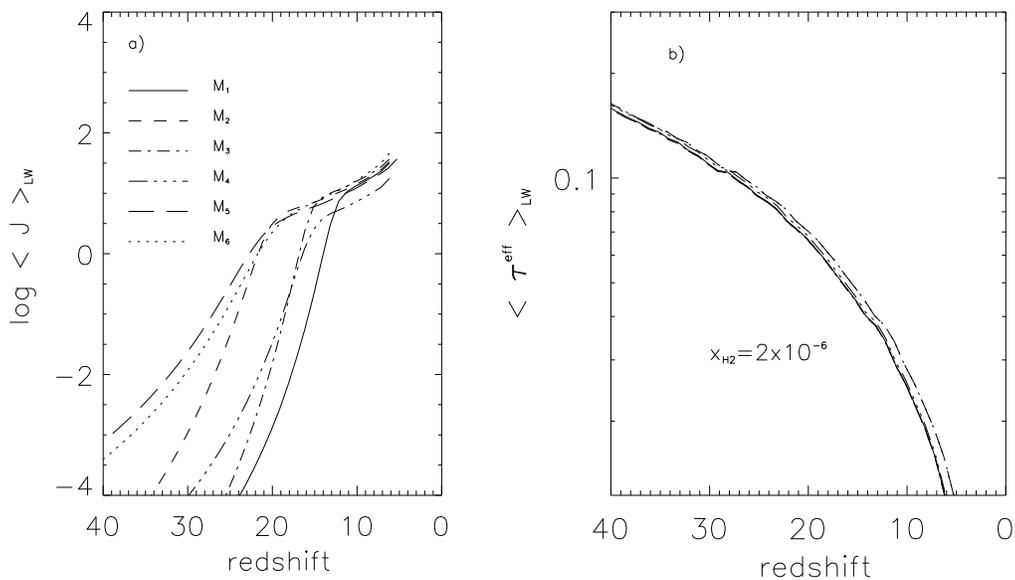}
\end{center}
\label{}
\end{figure}
Panel a) from Fig.~1 presents the evolution with  redshift
of the UV background flux averaged in  LW bands
obtained for different parameterizations of the comoving emission
coefficient as indicated in the Table.
Panel b) of the same figure presents the evolution with redshift
of the effective optical depth averaged in LW bands
computed by using the
same procedure as Haiman, Abel \& Rees (2000),
for a constant H$_2$ fraction of
2$\times$10$^{-6}$ ($\approx$ the H$_2$ fraction at recombination).
As at smaller redshifts the H$_2$ fraction
is substantially reduced, the effective optical depth
of the IGM due to absorption into  H$_{2}$ lines
is even smaller.
We found that by neglecting the absorption into H$_{2}$ lines
the UVB flux is affected by maximum 10\% for redshifts
higher than 30.\\
\begin{figure}
\caption{Left panel: Evolution with redshift
of the UVB flux at
$\nu^{th}_{H_{I}}=13.6$ eV ($\approx$1 Ryd).
The critical
UVB flux below which ($T_{vir}\leq 10^{2.4}$K)
 the star formation is prevented (thick dotted line)
is from Haiman, Abel \& Rees (2000).
Right panel: Redshift evolution of the
hydrogen photo-ionization rate for the UVB models presented
in the right panel. The observational upper limits  based on
Ly$\alpha$, Ly$\beta$ and Ly$\gamma$ Gunn-Peterson troughs at
 $z=6.05$  are from Fan et al. (2002).
We indicate by $M_5^{thin}$ the
hydrogen photo-ionization rate obtained
by neglecting the optical depth effects for model $M_5$.
The model parameters are indicated in the Table.}
\begin{center}
\includegraphics[width=15cm]{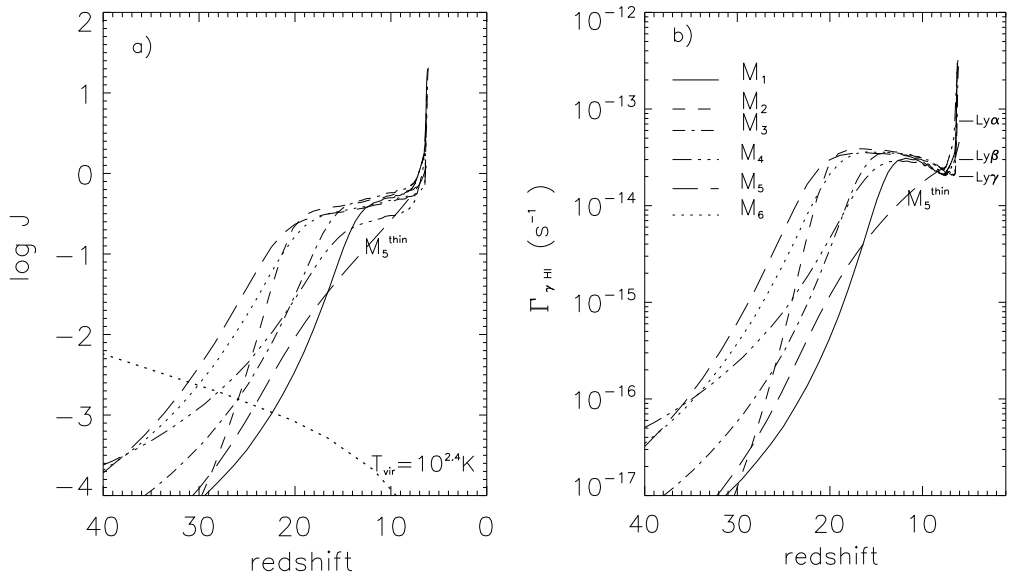}
\end{center}
\label{}
\end{figure}
Panel a) from Fig.~2  presents
the evolution with redshift
of the UVB flux at
$\nu^{th}_{H_{I}}=13.6$ eV ($\approx$1 Ryd).
We also indicate (thick dotted line) the critical
UVB flux below which ($T_{vir}\leq 10^{2.4}$K)
 the star formation is prevented
(Haiman, Abel \& Rees, 2000).

The photo-ionization rates, $\Gamma_{\gamma_{i} }$, and the photo-heating
rates, $\epsilon_{\gamma_{i}}$, of any species $i$
are related  to the UVB flux, $J(\nu,z)$, through:
\begin{eqnarray}
\Gamma_{\gamma_{i}}=\int_{\nu_{th_{i}}}^{\infty}
\frac{J(\nu,z) \sigma_{i}(\nu)}{h\nu}d\nu\,
\hspace{0.5cm}
\epsilon_{\gamma_{i}}=\int_{\nu_{th_{i}}}^{\infty}
\frac{J(\nu,z)\sigma_{i}(\nu)(h\nu-\nu_{th_{i}})} {h\nu} d\nu \,,
\end{eqnarray}
where $\sigma_{i}(\nu)$  are the photo-ionization (dissociation)
cross-sections (Abel et al. 1997)
and $\nu^{th}_i$ are the
ionizing threshold frequencies of  species $i$
($\nu^{th}_{H_{I}}$=13.6eV, $\nu^{th}_{He_{I}}$=24.6eV,
$\nu^{th}_{He_{II}}$=54.4eV).\\
We compute $\Gamma_{\gamma_{i} }$ and $\epsilon_{\gamma_{i}}$
for those choices of parameters
${\bf p}=({\rm A},{\rm B},z_{re},\alpha)$
in eq.~(4) that constraint the
hydrogen photo-ionization rate at $z \simeq 6$ to
$\Gamma_{\gamma_{HI}} \simeq  8 \times $10$^{-14}$
 s$^{-1}$, as indicated by the observational upper limits
  based on
 Ly$\alpha$, Ly$\beta$ and Ly$\gamma$ Gunn-Peterson troughs
 (Fan et al. 2002). \\
Panel b) from Fig.~2 shows the evolution with redshift of
hydrogen photo-ionization rates
corresponding to the UVB models presented in panel a).
We also indicate by $M_5^{thin}$
redshift evolution of the hydrogen photo-ionization rate obtained
by neglecting the optical depth effects for simulation $M_5$.
The comparison between  two cases
shows that the optical depth effects results
in a gradually increase of  photo-ionization
rate over a larger redshift interval.\\
\begin{figure}
\caption{
Redshift evolution of the
mean excess energy of \hi, \he and \hee ionizing photons
for simulation $M_1$ (solid line) and $M_2$ (dashed line).
}
\begin{center}
\includegraphics[width=10cm]{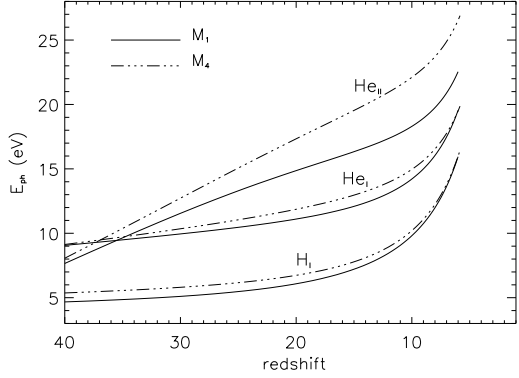}
\end{center}
\label{}
\end{figure}
The differences between various UVB models
can be understood in terms
of energy input of the ionizing photons
before and during the reionization.
In  Fig.~3 we show the redshift evolution
of the mean excess energy  of ionizing photons, $E^i_{ph}=\epsilon_{\gamma_{i}}/ \Gamma_{\gamma_{i}}$,
for  models $M_1$ and $M_4$.
Before and during reionization the  mean excess energy
is higher for  model $M_4$. For this case the redshift interval,
$\Delta z \sim B^{-1}$, in which the reionization is completed
is larger (see also Fig.~2).
For both models the mean excess energy is
dominated by the input energy
from  \hee ionizing photons. \\
We find that,  at $z \approx 6$
the mean excess energy from \hee ionizing photons
is $\sim$ 0.42$\nu^{th}_{He_{II}}$ for model $M_1$ and
$\sim$ 0.5$\nu^{th}_{He_{II}}$ for model $M_4$.
This corresponds to a \hee photo-heating rate about 1.2 times larger
for model $M_4$ at $z\approx 6$.
The increased  \hee photo-heating rate
results in an increase of
the IGM temperature before and during the reionization, as it has been demonstrated by Abel and Haehnelt (2000). \\

\section{The ionization fraction in the presence of UVB flux}

We modify the Smooth Particle Hydrodynamics (SPH)
code HYDRA(v4.2)\footnote{http://hydra.mcmaster.ca/hydra}
 (Couchman, Thomas and Pearce, 1995; Theuns et. al., 1998)
to study the redshift evolution of the  ionization fraction
in the presence of  UV radiation background
taking into account the radiative transfer effects,
as given by eq.~(1).

Cosmological hydrodynamic simulations have proven
successful in studying the IGM reionization at sub-galactic scales
(see e.g. Hernquist et al. 1996; Miralda-Escud\'e et al. 1996; Ciardi et al. 2000; Ciardi, Stoehr \& White, 2002).\\
It is useful to note the importance of the resolution and the volume
of simulations aimed the study of reionization process.
The resolution must be high enough to
follow the formation and
evolution of the objects responsabile
for producing the bulk of the ionization radiation.
At the same time,
a large simulation volume is required
in order to have a region with
representative properties and to avoid the bias
due to the variance on small scales.\\
Ciardi et al. (2000)  derived the minimum mass
of the objects that contribute substantially
to the reionization process. They show that at $z>15$
the main contribution to \hii  filling factor comes
from small mass objects, $M \sim 10^7 M_{\odot}$,
while at lower redshifts, when the formation of such objects
is suppressed by the feedback mechanisms, the main contribution
comes from objects with masses $M>10^9M_{\odot}$. \\
In Fig.~4 we show the mutual dependence
on the simulation box length
of the mass per gas particle, $M_b$,
and of the {\it rms} linear overdensity of a sphere
with the mass equal to that of the simulation box at present time,
$\sigma_0$.
One should note that once the value of $\sigma_0$ for a given run
is selected,  the redshift values  of simulation outpus
with $\sigma$ lower than $\sigma_0$
are uniquely specified [$\sigma \sim \sigma_0/(1+z)]$.

\begin{figure}
\caption{The mutual dependence on the simulation box length of the mass per gas particle, $M_b$,
(solid line) and of  the {\it rms}
linear overdensity of a sphere
with the mass equal to that of the simulation box at present time,
 $\sigma_0$, (dashed line).}
\begin{center}
\includegraphics[width=12cm]{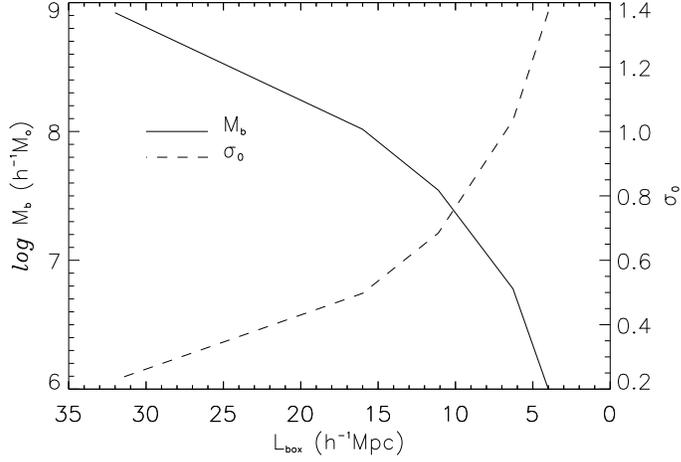}
\end{center}
\label{}
\end{figure}
In this paper we present cosmological
hydrodynamical simulations obtained for
$128^3$ dark matter  particles and the same number
of gas  particles  in boxes
with comoving lengths of 16$h^{-1}$~Mpc and
a total mass per cell of $M=6.1\times 10^8 h^{-1} M_{\odot}$
($M_b=1.02\times 10^8h^{-1}M_{\odot}$ for gas particle and
$M_{cdm}=5.1\times 10^{8}h^{-1}M_{\odot}$ for
cold dark matter particles).\\
This choice allows us to study large enough simulated regions and
to resolve objects with masses $M \sim 5 \times 10^8 M_{\odot}$,
responsable for  photon production at reionization
 redshifts $\sim$15, studied in this paper.
At the same time, this choice ensures  that at the end of the simulation
the  volume is large enough to provide a proper average
(e.g.  $\sigma \approx 0.07$ at $z=6$).\\
The initial particle positions and velocities
were set up at $z_i=50$ by using
the transfer functions
calculated with CMBFAST code (Seljak \& Zaldarriaga, 1996)
for our $\Lambda$CDM cosmological model.\\
As at early times  the Compton
cooling is very efficient
due to the coupling of the free electrons  (left from recombination)
to the CMB photons,
we assume an initial IGM temperature
$T \simeq T_{CMB}=2.726 \times (1+z) \simeq 139$K (Theuns et al. 1998).\\
Our cosmological simulations  include  the physical
processes relevant for the primordial gas dynamics in the
photo-ionized IGM
(Anninos et al. 1997; Abel et al. 1997;
Theuns et al. 1988 and references therein).
The baryonic gas is composed by neutral  (\hi, \he) and
ionic (\hii, \hee, \heee) species,
with a primordial composition of
76\% hydrogen and 24\% helium by mass.
The fractional densities of the various gas species are denoted with the
standard  nomenclature by
normalizing their comoving number densities to the present hydrogen
number density $n^0_{{\rm H}}$,
[ {\it e.g.}  \hi=$n_{{\rm H}_{{\rm I}}}/n^0_{{\rm H}}$,
 where $n^0_{{\rm H}}$=1.88$ \times 10^{-7}(\Omega_bh^2$/0.022) cm$^{-3}$].\\
The cooling model includes collisional ionization,
recombination, collisional excitation,
bremsstrahlung, inverse Compton cooling on
the CMB, and cosmological expansion cooling.
Also, the relevant heating mechanisms,
photo-ionization (dissociation) and photo-heating of hydrogen and helium
are  included. \\
HYDRA(v4.2) uses functional forms of cooling and recombination
rates
based on the atomic physics coefficients collected by Cen (1992)
for use in cosmological hydrodynamic simulations (see Appendix B from Theuns et al. 1988).
We check out them satisfactorily against those quoted
using  updated atomic data by Abel et al. (1997).\\
The simulations are integrated in time
up to a final epoch, $z_f$,
where the baryonic gas
reach its termal equilibrium determined by the balance between
the heating, $H_{ph}$, and cooling, $C$, rates  defined as:
\begin{eqnarray}
H_{ph}=(H_{I}\epsilon_{\gamma_{HI}} +
He_{I}\epsilon_{\gamma_{HeI}} +
He_{II}\epsilon_{\gamma_{HeII}})/n^0_H\,, \hspace{0.1cm}
C=\sum^{11}_{i=1}c_i(T)+C_{ad}(T)\,, \nonumber
\end{eqnarray}
where: $\epsilon_{ \gamma_{H_{I}} }$,
$\epsilon_{\gamma_{He_{I}}}$ and $\epsilon_{\gamma_{He_{II}}}$ are the
photo-heating rates of \hi, \he and \hee respectively and $c_i$
are the cooling rates whose functional dependence
on temperature are given in Appendix B1 from Theuns et al.(1998).\\
In the above equation  $C_{ad}=3k_B T H(z)\rho / \mu m_p $ is the adiabatic cooling
rate due to the cosmological expansion,  $H(z)$ is the Hubble expansion rate,
$k_B$ is the Boltzmann constant, $\mu$ is the mean molecular weight and
$m_p$ is the proton mass. \\
The code chooses the output time steps
determined by the maximum instantaneous
values of particle velocities and accelerations.
An optimal low-order integration scheme is used for
advancing particle positions and velocities (Couchman, Thomas, \& Pearce, 1995).\\
\begin{figure}
\caption{Bottom: Redshift evolution of the heating, $H_{ph}$, and cooling,
$C$, rates  (units of erg s$^{-1}$ cm$^{-3}$ )
for simulation $M^{thin}_5$ (thin solid and dashed lines)
and $M_5$ (thick solid and dashed lines).
 Top:  Coupled evolution of the IGM temperature $T$
 and ionization fraction $x_e$  obtained for
simulation $M^{thin}_5$ (thin solid and dashed lines)
and $M_5$ (thick solid and dashed lines).}
\begin{center}
\includegraphics[width=15cm]{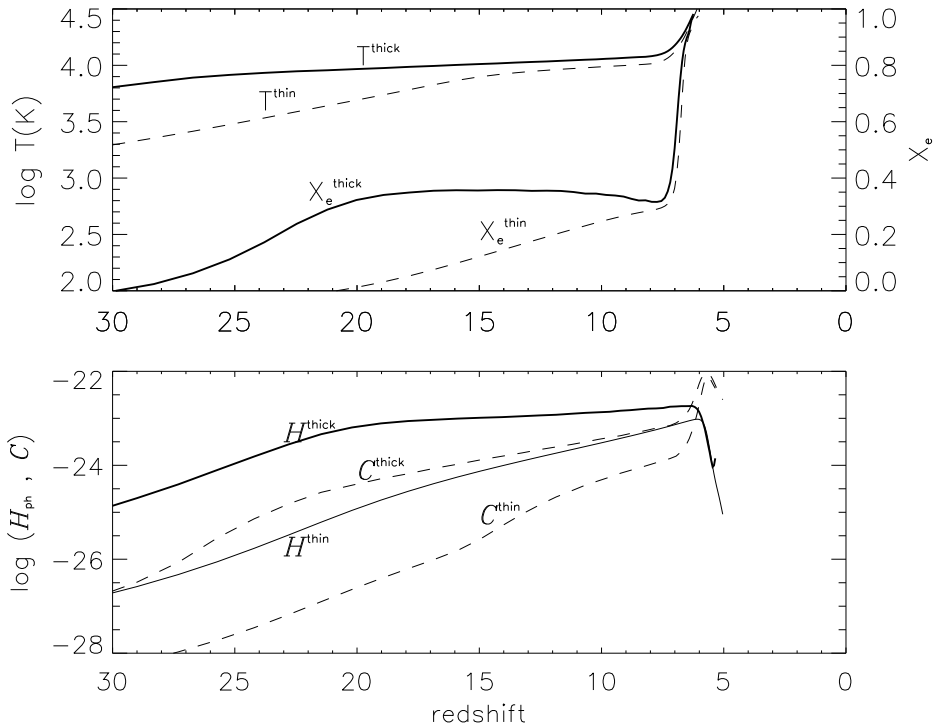}
\end{center}
\label{}
\end{figure}
At each simulation output time step, for each gas particle $i$,
we read out the density,
$\Delta_i=\rho_i/{\bar \rho}_b$ (in units of the mean baryon density,
${\bar \rho}_b$) and temperature, $T_i$,
(converted from the internal energy).
Then, we compute  the abundances of all
neutral and ionic species and the
ionization fraction defined  as:
\begin{equation}
 x_e(z)=\frac{e(z)} { e(z)+ {\rm H_{{\rm I}}}(z)+
 {\rm He_{{\rm I}}}(z)} \,,
\end{equation}
where e=\hii+\hee+\heee  is the total free electron fraction
(e=n$_e$/n$^0_{{\rm H}}$).
With this definition $x_e \rightarrow 1 $ for a fully ionized
gas and  $\chi \rightarrow 0 $ for a fully neutral gas.\\
Fig.~5 presents the redshift evolution of the
heating  and cooling rates (bottom panel) and
the coupled evolution of the IGM temperature
and ionization fraction (top panel) for
simulation $M_5$.
In order to stress the importance of the radiative transfer effects,
we present these dependences
obtained by neglecting the optcal depth  effects (thin case)
and by considering the optical depth effects (thick case).
The increase of photo-ionization and photo-heating rates
due to the  the optical depth effects
results in an increase of the IGM temperature,
leaving distinct features
on the temporal evolution of the ionization fraction.\\
In Fig.~6 we present a projection of the ionization fraction
in $\log \Delta_i$ - $T_i$ plane  at $z=6$ for
simulation $M_5$.\\
\begin{figure}
\caption{A projection of the ionization fraction
in $\log \Delta_i$ - $T_i$  plane  at $z=6$ from
simulation $M_5$.}
\begin{center}
\includegraphics[width=12cm]{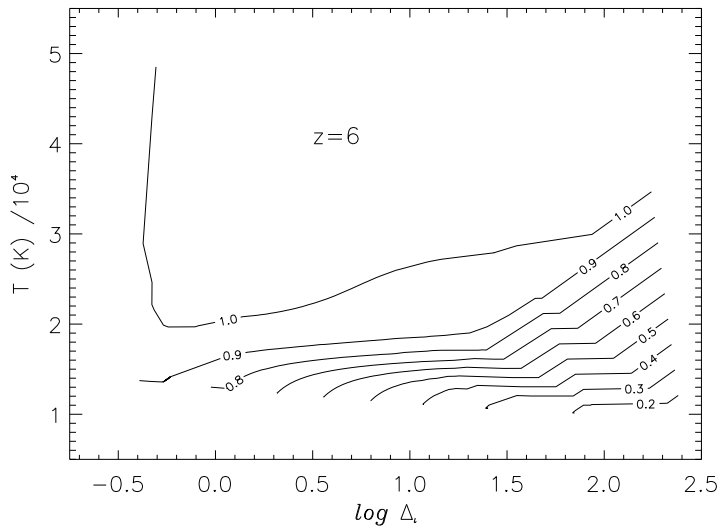}
\end{center}
\label{}
\end{figure}
The ionization histories obtained for
the UV radiation background  models
discussed before are presented in Fig.~7.
The model parameters are chosen
to investigate
the effects of Thomson optical depth in the range
$\tau_{es}$=0.05 - 0.1 and to study the possibility to
distinguish between models with the same values of $\tau_{es}$.
Models $M_{4}$ and $M_5$ are chosen to
have the same $\tau_{es}$ and the same value
of the reionization redshift (see also the Table).
We note that while different
ionization histories can be  distinguished by the
corresponding value of $\tau_{es}$, the same
value of $\tau_{es}$
can be produced by different parameterizations of UVB flux.\\
\begin{figure}
\caption{The evolution with redshift of the ionization fraction
obtained for the UVB models presented in Fig.~2.
The model parameters and the corresponding electron optical depth to the Thompson scattering, $\tau_{es}$, are indicated in the Table.}
\begin{center}
\includegraphics[width=12cm]{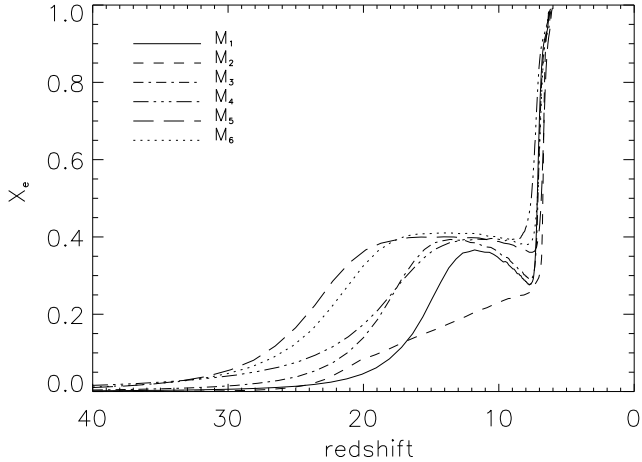}
\end{center}
\label{}
\end{figure}

\section{CMB angular power spectra in presence of the UV background}

To evaluate the effects of the
radiative feedback  on the
CMB angular power spectra
we  modified the CMBFAST code
to  include the  redshift evolution
of various ionization histories presented in Fig~6.
The corresponding CMB anisotropy temperature, $C_{T}(l)$, and
E-mode polarization, $C_{E}(l)$,
 power spectra are presented in Fig.~8. \\
One should note in Fig.~8 that while $C_{T}(l)$
power spectra are almost degenerated,
the differences in different reionization histories
produce undegenerated signatures on $C_{E}(l)$
power spectra at low multipoles $(l\leq 50)$.
This can be explained by the fact that while polarization
is projecting from the epoch of reionization
at angular frequencies $l=k(\eta_0-\eta_{ri})$
( here $k$ is the wave number, $\eta_0$ and $\eta_{ri}$
are the conformal times today and at the epoch of reionization)
 the temperature is projecting from the (further)
last scattering surface (Zaldarriaga, 1997). \\
\begin{figure}
\caption{The CMB angular power spectra, $C_{T}(l)$ and
$C_{E}(l)$ for  the reionization histories
presented in Fig.~7. The model parameters and
the corresponding electron optical depth
to the Thompson scattering, $\tau_{es}$,
are indicated in the Table. }
\begin{center}
\includegraphics[width=14cm]{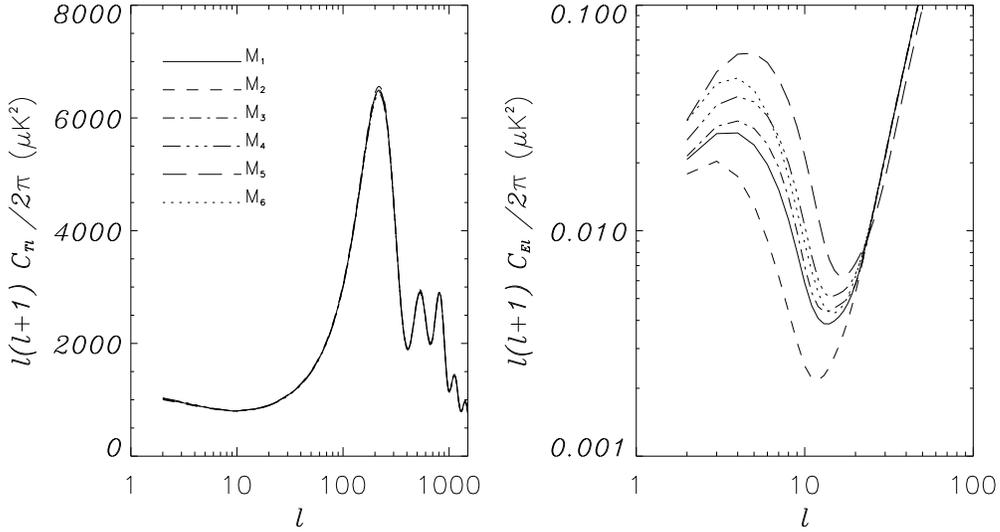}
\end{center}
\label{}
\end{figure}
The differences between the reionization models in comparison with the
expected sensitivity of the future {\sc Planck} mission
can be expressed in terms of relative difference
between the power spectra $C_{E}(l)$
of the E polarization component as (Naselsky and Chiang, 2004):
\begin{equation}
D_{i,j}(l)=\frac{2[C_{E,i}(l)-C_{E,j}(l)]}{C_{E,i}(l)+C_{E,j}(l)} \,,
\end{equation}
where the indices $i$ and $j$ denote different reionization models.
We compare the amplitude of the function $D_{i,j}(l)$ with
the expected relative error of the $C_{E}(l)$ anisotropy power spectrum for
the  {\sc Planck} experiment. If  systematic  and foreground
effects
are successfully removed, the corresponding error bar is given by:
\begin{equation}
\frac{\Delta C_E(l)}{C_E(l)}=\frac{1}{\sqrt{f_{sky}(l+\frac{1}{2})}}
[1+w^{-1}C^{-1}_E(l)W^{-2}_l)] \, ,
\end{equation}
where $f_{sky} \simeq$ 0.65 is the sky coverage,
$w=(\sigma_p\theta_{FWHM})^{-2}$, $\sigma_p$ is the sensitivity
per resolution element $\theta_{FWHM} \times \theta_{FWHM}$,
$W_l={\rm exp}[-l(l+1)/2 l^2_s]$ is the beam window function and
$l_s=\sqrt{2{\rm ln}2}\theta^{-1}_{FWHM}$.
For all the frequency channels of the {\sc Planck} experiment
$\theta_{FWHM}$ is less then
$\sim 30$ arcminutes and at  low multipoles ($l \lsim 50$)
 the dominant term in eq.~(8) is the first term,
due to the combined cosmic and sampling variance.\\
\begin{figure}
\caption{Relative differences between couples of
E-mode polarization angular power spectra
corresponding to models having the same value of $\tau_{es}$:
$D_{12}$ for models $M_1$ and $M_2$ (solid line),
$D_{34}$  for models $M_3$ and $M_4$ (dashed line) and
$D_{56}$  for models $M_5$ and $M_6$ (dash-dotted line).
The model parameters and
the corresponding electron optical depth
to the Thompson scattering, $\tau_{es}$,
are indicated in the Table. }
\begin{center}
\includegraphics[width=13cm]{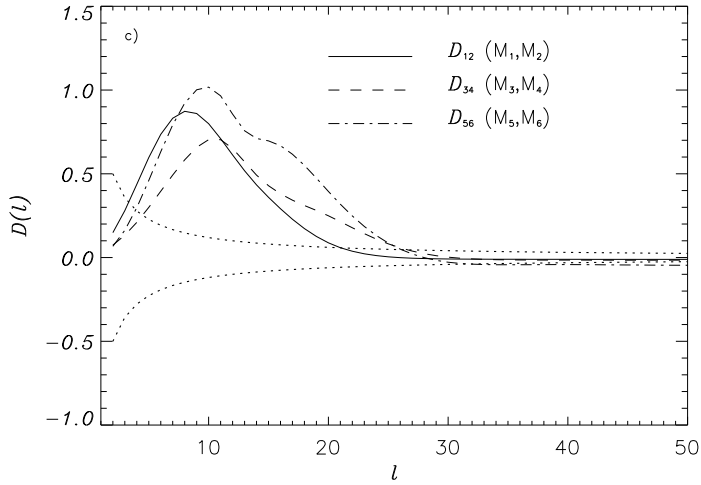}
\end{center}
\label{}
\end{figure}
In Fig.~9 we plot the function $D_{i,j}(l)$ obtained for the
models with the same value for $\tau_{es}$
compared with the  {\sc Planck} sensitivity  at the
corresponding multipoles.
One can see that in all cases the deviation function
$D_{i,j}(l)$ lies above the cosmic variance region,
indicating that the upcoming {\sc Planck} satellite will allow to
distinguish among E-polarization power spectra with the same
electron optical depth $\tau_{es}$
and also the same reionization redshift $z_{re}$, provided that
they have been generated by sufficiently different
reionization histories.\\
We observe that, in practice, the possibility to
determine the CMB angular power spectra at $l \lsim 50$ with a sensitivity
limited essentially only by cosmic and sampling variance
relies on the accuracy of the removal of systematic effects
(see, e.g., the review by Mennella et al., 2004 and reference therein)
and foreground contamination. In particular, even for an ideal
experiment (i.e. without systematic effects) Galactic
foregrounds represent the main limitation in the recovery of the CMB
angular power spectra. The accuracy of their modeling and subtraction
from microwave maps depends on the experiment sensitivity and resolution.
This aspect is particularly crucial for CMB polarization
anisotropy analyses,
because of the low polarization degree of the CMB anisotropy.
Baccigalupi et al. (2004) show that at the {\sc Planck}
sensitivity level a blind (i.e. not critically dependent on the
assumptions on the properties of the maps to be recovered)
and accurate (i.e., in this context, precise enough to avoid
significant errors in the recovery of the angular power spectrum
of the CMB component) separation between the Galactic
and CMB polarized signals from multifrequency maps
is possible at multipoles
$l \lsim 100$ while a worse sensitivity has
a critical impact on the accuracy of this separation.
For this reason
a high sensitivity experiment, like {\sc Planck},
is crucial to remove all systematic effects and separate
the CMB from the foregrounds with an accuracy good enough
to measure the CMB E-mode signal at the limit of the
cosmic variance.

\section{Conclusions}

In this paper we study the role of radiative  effects
 for the reionization
history of the universe and address the
detectability
of their features
by the future CMB polarization measurements.\\
We present cosmological hydrodynamical simulations
at sub-galactic scales that allow
to resolve objects with masses of $M \sim 5 \times 10^8 M_{\odot}$,
responsable for the photon production at
reionization redshifts $z_{re} \sim 15$,
ensuring in the same time
a proper average of the simulation
the  volume  at smaller redshifts.\\
Our  cosmological simulations include
the radiative mechanisms relevant for the primordial
gas dynamics: photo-ionization, photo-heating  and cooling of the
hydrogen and helium in the expanding universe.\\
We compute the mean specific UVB flux as  solution
to the radiative transfer equation
for those parameterizations of the
comoving emission coefficient that
constaine the hydrogen photo-ionization rate value
to the upper limit indicated by the experimental measurements.\\
We find that the increase of photo-ionization and photo-heating
rate due to optical depth effects results in a
significantly contribute to the heating of the IGM
before and during the reionization. \\
We show that the main effect of the
UV radiation spectrum
on the temporal evolution of the ionization
fraction is given by the value of the reionization
redshift, $z_{re}$, and the redshift interval,
$\Delta z$, in which the reionization is completed.
As reionization proceeds, the excess energy
of ionizing photons build up a sufficient UV radiation background
that rises the IGM temperature.
The net effect is  a decrease of the global
ionization fraction with cosmic
time over a limited period. \\
We evaluate the effects of the UV radiation background
on the CMB angular power spectrum
taking into account different
temporal evolutions of the ionization fraction.\\
We find that for reionization models
with degenerated CMB temperature anisotropy power spectra, $C_T$,
the radiative feedback mechanisms leave
distinct signatures on the E-mode polarization power spectrum, $C_E$,
at large scales ($l<50$).
We  show that through E-mode CMB polarization
power spectrum measurements,
the {\sc Planck} experiment will have the sensitivity to distinguish
between
different reionization histories even when
they imply the same optical
depth to electron scattering and
degenerated $C_T$ power spectra.

\section{Acknowledgments}

We thank to reviewers for the helpful comments.
We acknowledge the use of the computing system at
{\sc Planck}-LFI Data Processing Center in Trieste
and the staff working there.
We thank to M. Sandri for useful suggestions.

\newpage
\begin{table}[]
\caption[] {Parametrization of the comoving emission coefficient
for the UVB flux models  studied in this papers.
We also indicate the corresponding electron
optical depths to the Thomson scattering,$\tau_{es}$.}

\vspace{0.5cm}
\begin{small}
\begin{center}
\begin{tabular}{ccc}
\hline \hline
Model  & ${\bf p}=({\rm A},{\rm B},z_{re},\alpha)$ & $\tau_{es}$   \\ \hline
M$_{1}$& (0.065, 11.25, 12, 1.8) & 0.05 \\
M$_{2}$& (0.068, 12.50, 17, 2) & 0.05 \\
M$_{3}$& (0.075, 13.75, 15, 2) & 0.08 \\
M$_{4}$& (0.039, 10.00, 15, 1)     & 0.08 \\
M$_{5}$& (0.057,  7.70, 18, 1.8)& 0.1 \\
M$^{thin}_{5}$& (0.012,  7.70, 18, 1.8)& 0.04 \\
M$_{6}$& (0.050, 6.90, 17, 1.8  )& 0.1 \\ \hline
\end{tabular}
\end{center}
\end{small}
\end{table}

\end{document}